\documentstyle[12pt,twoside,fleqn,espcrc1,psfig]{article}
\def\slap#1#2{\setbox0=\hbox{$#1{#2}$}
        #2\kern-\wd0{\hbox to\wd0{\hfil$#1{/}$\hfil}}}
\def\feyn#1{\mathpalette\slap{#1}}
\title{The Phase Diagram of QCD}
\author{M.\,A. Stephanov\address{Institute for Theoretical Physics, 
SUNY, Stony Brook, NY 11794-3840, USA}}

\begin{document}
\maketitle

\begin{abstract}
We show that current experimental knowledge of QCD together
with general model independent arguments such as continuity, universality
and thermodynamic relations, as well as the information gained from
various models can be used to constrain the phase diagram of QCD
as a function of temperature and baryon chemical potential.
\end{abstract}

\section{Introduction}

Understanding the phase diagram of QCD as a function of temperature,
$T$, and chemical potential of the baryon charge, $\mu$, is an
important ingredient in the analysis of the results of heavy ion
collision experiments. We know a good deal about the behavior of QCD
at finite temperature. The basis for our understanding is provided by
the first principle lattice QCD Monte Carlo calculations. In
particular, we know that QCD with two massless quarks undergoes a
phase transition to the quark-gluon plasma phase at a temperature of
about 160 MeV.  In contrast to that our understanding of
the behavior of QCD at finite baryon charge density, or finite $\mu$,
is extremely poor. This is due to the unfortunate fact that lattice
QCD calculations based on Monte Carlo methods are not possible because
the measure of the Euclidean path integral is complex when $\mu$ is
not zero. In this report, based on the work ~\cite{paper}, 
we put together available experimental facts about QCD, 
results from various models, and apply
general model independent arguments such as continuity, universality
and thermodynamic relations in order to construct the phase diagram of
QCD in the $T\mu$ plane.

We perform our analysis for the two-flavor QCD --- a well-known and
phenomenologically successful approximation to real QCD. The effects
of the strange quark and electroweak interactions lead to
quantitative and in some cases qualitative modifications which we also
consider. The theory is described by a partition function:
\begin{equation}\label{z}
Z \equiv e^{-\Omega(T,\mu)/T} = \int {\cal D}A {\cal D}\bar\psi {\cal D}\psi
\exp\{-S_E\} \ .
\end{equation}
The Euclidean action, $S_E$, is given by
\begin{equation}
S_E = \int_0^{1/T} dx_0 \int d^3x 
\left[ {1\over2g^2} {\rm Tr} F_{\mu\nu}F_{\mu\nu} 
- \sum_{f=1}^{N_f} \bar\psi_f\left(\feyn\partial + \feyn{A}
+ m_f + {\mu\over N_c}
\gamma_0\right)\psi_f
\right] \ ,
\end{equation}
where $N_f=2$ is the number of flavors, $N_c=3$ is the number of
colors, and $m_f=m=0$ is the quark mass.
The Euclidean matrices $\gamma_\mu$ are hermitian. 
The normalization of $\mu$ differs from the normalization 
customary in lattice calculations by a factor $1/N_c$ (i.e., the baryon 
charge of a quark).

What do we actually mean by understanding the phase diagram?
The most prominent features of a phase diagram are phase transitions.
They manifest themselves through the singularities or discontinuities
in the dependence of various thermodynamic observables on the
parameters $T$ and $\mu$. Such observables can be obtained by
differentiating the thermodynamic potential $\Omega$ with respect to
$T$ and $\mu$:
\begin{equation}\label{ns}
nV= \sum_f\langle\bar\psi_f\gamma_0\psi_f \rangle
= -{\partial\Omega\over \partial\mu}; \qquad
sV = -{\partial\Omega\over \partial T};
\quad \mbox{and also} \quad
\langle\bar\psi\psi\rangle N_f V
= -{\partial\Omega\over \partial m};
\end{equation}
where $n$ is the baryon number density, and
$s$ is the entropy density. These are the densities of extensive
quantities, such as baryon charge and entropy, per volume, $V$. 
The pressure, 
$pV=-\partial\Omega/\partial V=-\Omega$, is not independent
from $T$ and $\mu$ (and also $m$):
\begin{equation}
dp = sdT + nd\mu + \langle\bar\psi\psi\rangle N_f dm.
\end{equation}
This equation can be used to derive Clapeyron-Clausius-type
relations between the slopes of the first-order transition
lines and the discontinuities of $s$, $n$ 
and $\langle\bar\psi\psi\rangle$ \cite{paper,leutwyller,italians}.

There are two thermodynamic observables which turn out to be more
useful than others in discovering phase transitions: $n$ and
$\langle\bar\psi\psi\rangle$. This is because both are good order
parameters, i.e., they vanish {\em identically} in one phase and are
nonzero in the other. There must be a singularity, and thus
a phase transition, separating such two phases.
Theoretically, we understand, at least qualitatively, the behavior
of $\langle\bar\psi\psi\rangle$ because it is an order parameter of
a global symmetry, SU(2)$_L\times$SU(2)$_R$, and it distinguishes
two phases with two different realizations of this symmetry:
spontaneously broken and exact. However, phenomenologically
we know little about this phase boundary (yet!).
On the contrary, it is harder to
understand theoretically the behavior of $n$, but, fortunately,
we happen to live near the phase boundary separating phases with
$n=0$ and $n\not=0$. As a result, we have a good empirical and 
quantitative knowledge about this phase transition.

\section{Zero $T$}

To see why $n$ can serve as a good order parameter consider
the partition function (\ref{z}) written in the form of the Gibbs
sum over the quantum states of the system characterized by their
energy, $E$, and baryon charge, $N$:
\begin{equation}\label{gibbs}
Z = \sum_\alpha \exp\left\{-{E_\alpha - \mu N_\alpha\over T}\right\}\ .
\end{equation}
In the limit $T \to 0$, the state with 
the lowest value of $E_\alpha-\mu N_\alpha$ determines the properties
of the system. This is the ground
state at given $\mu$. Let us introduce
\begin{equation}
\mu_0 \equiv \min_\alpha (E_\alpha/N_\alpha) \ .
\end{equation}
As long as $\mu < \mu_0$ no state with nonzero $N$
can compete with the vacuum state ($E=0$, $N=0$) for the role of 
the ground state. Therefore, as long as $T=0$ and $\mu < \mu_0$, 
the equation $n(\mu) = 0$ holds exactly.
If $n\not=0$ when $\mu > \mu_0$, the point $\mu=\mu_0$ must
be a singular point. 

\begin{figure}
\centerline{\psfig{silent=,file=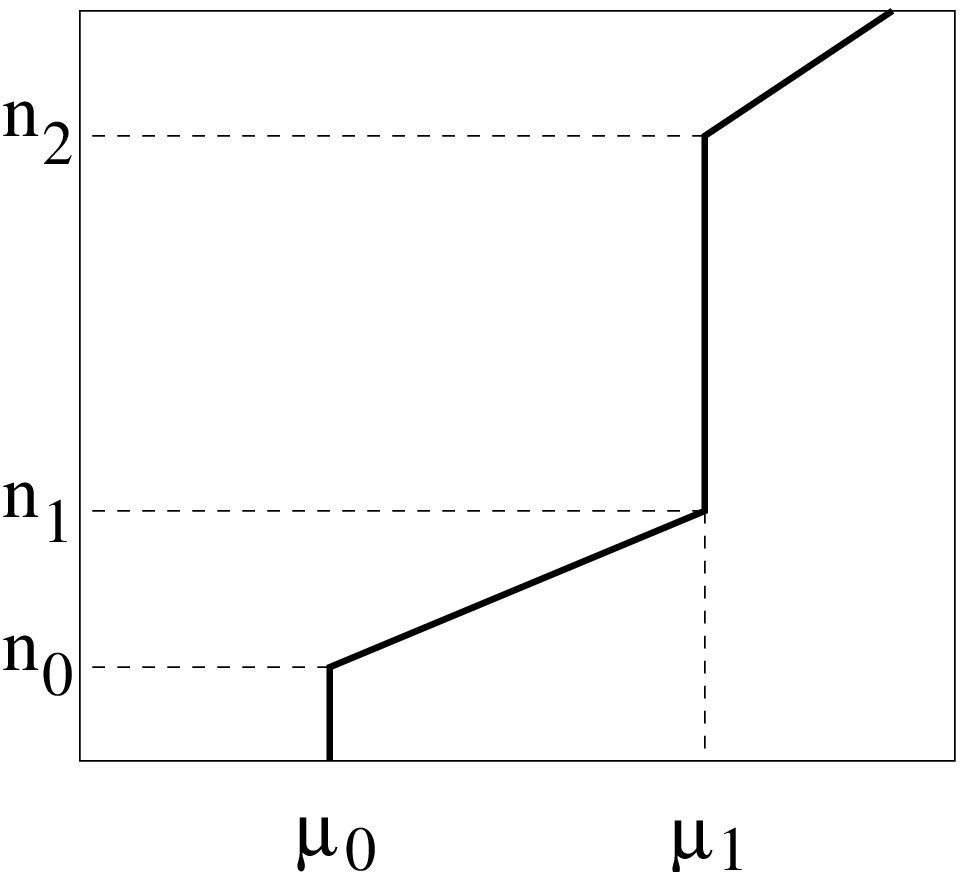,height=1.8in}\qquad\qquad
\psfig{silent=,file=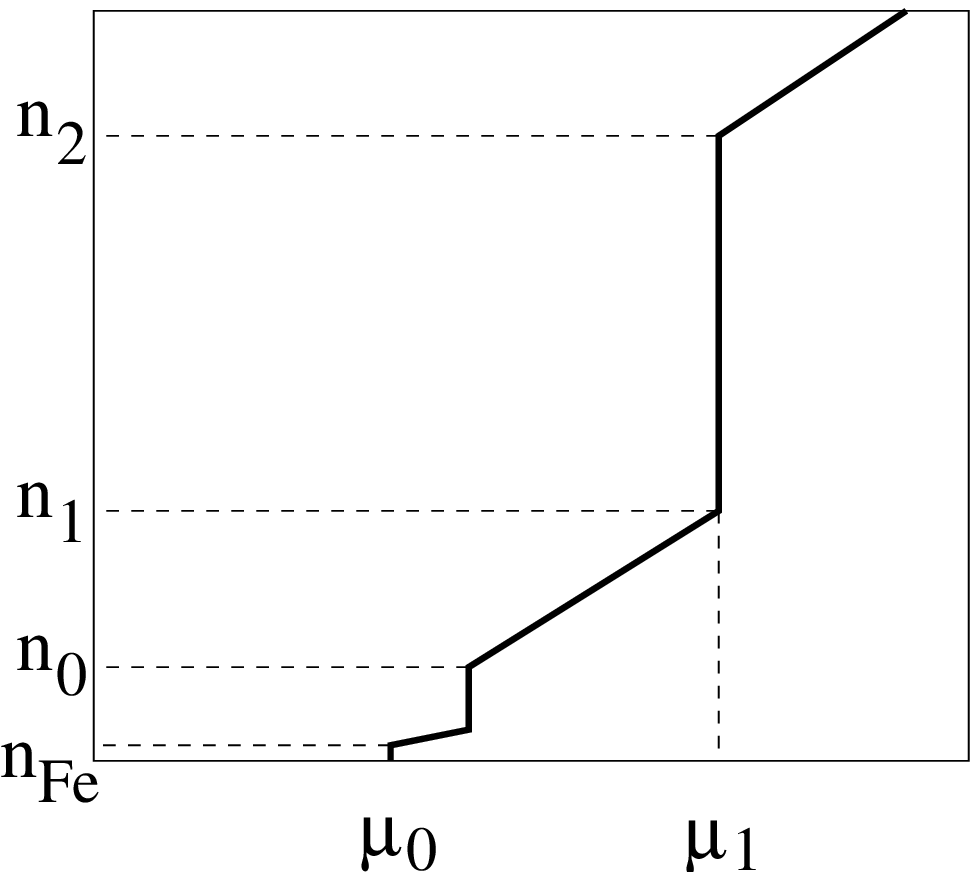,height=1.8in}\\}
\centerline{\hss (a) \hss (b) \hss}
\caption[]{Schematic dependence of the baryon charge density
on the chemical potential at $T=0$ (a) in QCD ($\mu_0\approx m_N-16$ MeV) 
and (b) in QCD+ ($\mu_0\approx m_N-8$ MeV). At $\mu_1$ the chiral
symmetry restoration transition occurs.}
\label{fig:nmu}
\end{figure}

In QCD without electromagnetism 
this singularity is a first order phase
transition separating vacuum phase from nuclear matter phase,
distinguished (at $T=0$ only) by the order parameter $n$.
The function $n(\mu)$ has a step at 
$\mu=\mu_0\approx m_N -16$ MeV equal to the density
of nuclear matter $n_0\approx 0.16\,{\rm fm}^{-3}$  at zero pressure as in
Fig.~\ref{fig:nmu}a. The slope of $n(\mu)$ just above $\mu_0$ is also known.
In the real world, the electromagnetic interaction and the presence
of electrons produce a tiny step at 
$\mu=\mu_0\approx m_N - 8$ MeV to the density of iron.
The step to the density of neutron matter occurs at 
somewhat larger value of $\mu$ (see
Fig. ~\ref{fig:nmu}b) at non-zero pressure  (as in neutron stars).

Increasing $\mu$ further takes QCD into the region about which we
have very little reliable theoretical or experimental
information. Various interesting phenomena have been predicted using
different models. Rigorously, however, one can only expect that the function
$n(\mu)$ continues to grow, to satisfy thermodynamic stability. Since
$n$ is non-zero it cannot serve as a good order parameter anymore and
we shall turn to another one: $\langle\bar\psi\psi\rangle$.

At very large $\mu$ the ground state of the system is to a good
approximation a Fermi sea of quarks. (This fact also leads to an
interesting possibility of quark-quark pairing around the Fermi
surface and color superconductivity~\cite{colorsc}.)  Thanks to the
asymptotic freedom and the screening of color interactions by the sea
of quarks, nonperturbative effects are suppressed. This motivates the
conclusion that at very large $\mu$ the condensate
$\langle\bar\psi\psi\rangle$ vanishes (provided that quark masses are
zero). If we denote by $\mu_1$ the value of $\mu$ such that
$\langle\bar\psi\psi\rangle=0$ for $\mu>\mu_1$ and
$\langle\bar\psi\psi\rangle\ne0$ otherwise, then the point 
$\mu=\mu_1$ must be a singular point.  It separates two phases with
distinct realizations of the global SU(2)$_L\times$SU(2)$_R$ chiral
symmetry.

There has been a wealth of theoretical research on the phase
transition at $\mu=\mu_1$ (see, e.g.~\cite{klevansky}). However, since
no first principle lattice calculations are possible at present, the
best we have is a collection of estimates obtained in various models
approximating the behavior of QCD at the chiral symmetry restoration
transition. The common denominator seems to be that the transition is
most likely of the first order. The value of $\mu_1$ is
somewhere of the order of 1 GeV.  Also, the empirical fact that stable
nuclear matter with $n\ne0$ and $\langle\bar\psi\psi\rangle\ne0$
exists indicates that $\mu_1>\mu_0$.
(The strange quark
tends to lower the value of $\mu_1$. An interesting, but empirically
disfavored, possibility that $\mu_1<\mu_0$, strange quark matter
at zero pressure, arises in this case~\cite{strange-matter}.)

\section{Nonzero $T$}

At finite $T$ the baryon density $n$ is no longer a good order
parameter. However, since the transition at $\mu_0$ is of the
first order, continuity ensures that it remains first-order
for sufficiently small $T$. The end-point of this
transition is a critical point in the Ising
universality class, which is probed in the multifragmentation
experiments~\cite{multifrag}.

The chiral condensate $\langle\bar\psi\psi\rangle$ is a good
order parameter at $T\ne0$, 
as long as $m=0$. This means that the
regions $\langle\bar\psi\psi\rangle\ne0$ at low $T$ and $\mu$ and
$\langle\bar\psi\psi\rangle=0$ at high $T$ and $\mu$ must be separated
by a phase transition. At $\mu=0$ lattice simulations predict that the
chiral symmetry is restored at $T_c\approx 160$ MeV for two-flavor
QCD~\cite{TC93}. A beautiful argument~\cite{PiWi} suggests that this
transition is likely to be a second order transition in the
universality class of $O(4)$ spin models in 3 dimensions.

At nonzero $\mu$ there is a line of second order phase
transitions in the $O(4)$ universality class
(the $\mu$ direction is not relevant near this critical
point). Since this line cannot terminate, and since at $T=0$
the transition is, presumably, of the first order, a logical
possibility arises that the transition turns first order in some
point $T_3$, $\mu_3$ on the phase diagram. This point has been
observed in various models of QCD chiral phase 
transition~\cite{italians,klevansky,paper,BR}.
We wish to point out that the critical behavior near this
point is determined by universality (the observation
also made independently in \cite{BR}). Study of tricritical 
points shows~\cite{lawrie} that the upper critical
dimensionality for such a point is 3 and, therefore, in QCD 
with two massless quarks the critical
behavior near the tricritical point must be given by the mean field
theory up to logarithmic corrections. 

In this work we used a
random matrix model to describe the chiral phase transition.
In accordance with generic expectations and other models 
it predicts a tricritical point.

\section{A random matrix model at finite $T$ and $\mu$}

A successful and very simple model which describes the degrees of freedom
of QCD related to the spontaneous breaking of chiral symmetry is the
random matrix model. It is based on the famous observation of Banks
and Casher that the value of $\langle\bar\psi\psi\rangle$ is related
to the density of small eigenvalues of the Dirac operator $\rho_{\rm
ev}(0)$.  Since we do not need to describe all the degrees of freedom
of QCD, but only those relevant to chiral symmetry breaking, a natural
and simple approximation of the Dirac operator by a random matrix
arises~\cite{rmt}. This approach is reminiscent of the one introduced
by Wigner in the study of spectra of heavy nuclei.  The power of the
random matrix model in describing the chiral symmetry breaking and
restoration at finite chemical potential is in the fact that, on
the one hand, this model shares an important property of QCD at
$\mu\ne0$ --- the complex fermion determinant, and it is exactly
solvable, on the other hand. A successful example is the explicit
demonstration that at nonzero $\mu$ quenched QCD is not a 
smooth $N_f\to0$ limit of real QCD~\cite{St96}.

\begin{figure}
\setlength{\unitlength}{2.4in}
\centerline{\psfig{silent=,file=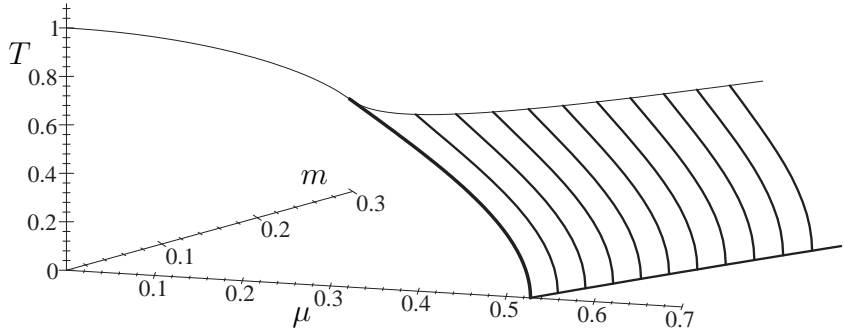,width=2\unitlength}}
\vspace{-\unitlength}
\begin{picture}(2,1)
\put(.34,.62){ $T$}
\put(.96,.06){ $\mu$}
\put(1.00,.36){$m$}
\end{picture}
\caption[]{Phase diagram of QCD with two light flavors
of mass $m$ as calculated from the random matrix model. 
The almost parallel curves on the wing surface
are cross sections of this surface with $m=$const planes.
The units of $m$ are $100$ MeV, of $T$ are
$T_c\approx 160$ MeV, of $\mu$ are $\mu_1/0.53\approx 2300$ MeV,
with the choices of $T_c$ and $\mu_1$ from the text.
}
\label{fig:pdrmt}
\end{figure}

The phase diagram calculated in the random matrix model~\cite{paper}
is shown in Fig.~\ref{fig:pdrmt}. We see that the second order phase
transition line in the plane $m=0$ turns into a first order line at
the tricritical point. The coordinates of this point are given by:
${T_3/ T_c} \approx 0.78$,  ${\mu_3/\mu_1} \approx 0.61$.
Taking $T_c=160$ MeV and $\mu_1=1200$ MeV, we find that $T_3 \approx 
120$ MeV and $\mu_3\approx 700$ MeV. A similar estimate has been
obtained recently using a different model~\cite{BR}.

\section{Conclusions}

\begin{figure}
\centerline{\psfig{silent=,file=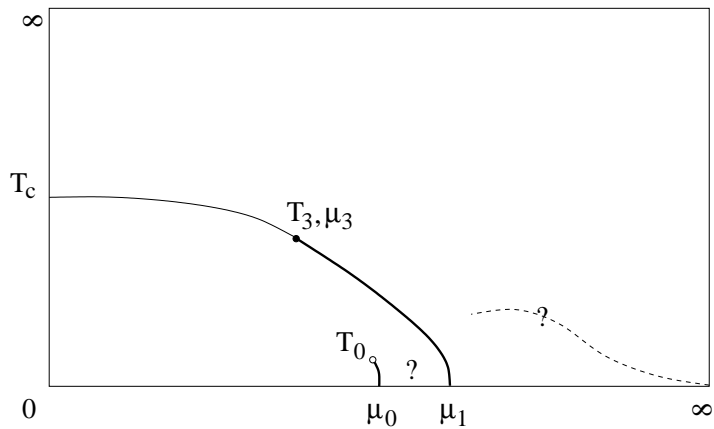,height=2.2in}}
\caption[]{A schematic phase diagram of QCD with 2 massless
quark flavors. Other phase transition lines are possible, for example,
in the low temperature region to the right of $\mu_0$. Another
example is a transition associated with color superconductivity
plotted as a dashed line. Thicker lines are first-order phase
transitions. The $T_c-T_3$ line is a second-order phase transition.
The tricritical point is at $T_3$, $\mu_3$  and the critical point
of the nuclear matter liquid-gas transition is at $T_0$.}
\label{fig:pd}
\end{figure}

We conclude with a sketch of the phase diagram (Fig.~\ref{fig:pd}) of
QCD with two massless quarks which
we find by analyzing the behavior of two thermodynamic quantities:
$n$ and $\langle\bar\psi\psi\rangle$. These quantities are
distinguished by the fact that both are good order parameters in
a certain sense: they identically vanish in some region of the phase
diagram and are non-zero in the other. The change of the behavior
of such a parameter from one region to the other is qualitative
and must proceed through a thermodynamic singularity.

Perhaps, the most interesting feature of this phase diagram is the
presence of the tricritical point. At small nonzero quark mass the
main change in the phase diagram is the disappearance of the second
order phase transition line (see Fig.~\ref{fig:pdrmt}). 
The first-order line remains, but it no longer separates
phases with different symmetry properties --- the chiral symmetry is
explicitly broken.  What is important is that the criticality at the
end-point of the first-order phase transition line remains.

The strategy for locating this end-point in the heavy ion collision
experiments is discussed in~\cite{srs}.  The signatures proposed
in~\cite{srs} are based on universal thermodynamic properties of the
critical point. For example, the divergence of the heat capacity will
lead to suppression of the event-by-event fluctuations of the apparent
temperature. Another signature is due to the long-wavelength
fluctuations of the sigma field near the critical point, which leads
to enhanced production of soft pions. An inspiring example that the
study of critical behavior in heavy ion collisions may, in principle,
be possible is provided by the multifragmentation
experiments~\cite{multifrag}, which can probe the end-point of the
nuclear liquid-gas phase transition.

\end{document}